\documentstyle[aps,twocolumn,epsfig]{revtex}

\begin{document}
\wideabs{
\title{Transverse energy fluctuations and the pattern of J/$\psi$
suppression
in Pb--Pb collisions}
\author{Jean-Paul BLAIZOT, Phuong Mai DINH, Jean-Yves OLLITRAULT}
\address{Service de Physique Th\'eorique, CE-Saclay\\
F-91191 Gif-sur-Yvette cedex}
\date{\today}
\maketitle

\begin{abstract}
The NA50 collaboration has recently observed that   the  J/$\psi$
production rate in Pb--Pb collisions decreases more rapidly as a
function of the transverse energy for the most central
collisions than  for less central ones. We show that this phenomenon can be
understood as an effect of transverse energy fluctuations
in central collisions.  A good fit
of the data is obtained  using a model which
relates $J/\psi$ suppression to the local energy density. 
Our results suggest that the $J/\psi$ is completely suppressed at 
the highest densities achieved in Pb--Pb collisions.
\end{abstract}
\pacs{25.75.-q 25.75.Dw}
}

Among the various particles which are produced in nucleus-nucleus
collisions at high energy, the $J/\psi$ meson plays a special role.
Because of the large mass of the charm quark, $c\bar c$ pairs are produced
on  a short time scale and their evolution probes the state of matter
in the early stages of the collisions. In particular, their
binding into  $J/\psi$ mesons may be  hindered by the presence of a
quark-gluon plasma in the collision zone where they are
produced, leading to the so-called 
``$J/\psi$ suppression'' \cite{Matsui86}. Indeed,  the rate of $J/\psi$
production in nucleus-nucleus collisions is smaller than what can be
expected on the basis of extrapolations based on independent
nucleon-nucleon collisions. However, the 
effect observed in collisions involving oxygen and sulphur
projectiles can be attributed to the so-called nuclear
absorption, a mechanism also at work in proton-nucleus collisions
\cite{Capella88,GH88,Nardi97}. 
But nuclear absorption alone is insufficient to account for the
large suppression observed in Pb--Pb
collisions, which has been qualified for this reason as
``anomalous suppression''  \cite{NA50-96}. Further final state
interactions seem to be needed to explain the data, and various scenarii
based either on hadronic interactions \cite{Hadronic},
or quark-gluon plasma
formation \cite{QGP,BLAIZ-OLLI}
have been proposed with various degrees of success \cite{VOGT}. 

The latest data obtained by the NA50 collaboration \cite{NA50-2000} allow
further progress. These data
  provide  new information on the dependence of the $J/\psi$ production
rate on the total transverse energy ($E_T$). In particular,  it is found
that the rate of $J/\psi$ production decreases faster
with increasing $E_T$ in the most central Pb--Pb collisions 
than in the less central ones. It has been suggested
that the resulting pattern of
$J/\psi$ suppression as a function of $E_T$ exhibits the two drops
corresponding  to the successive meltings of charmonium bound states,
first the $\chi$ 
(which is expected to produce 40\% of observed $J/\psi$)
and then the $J/\psi$ \cite{NA50-2000,Gupta92}. 
We shall in fact argue that the extra suppression
observed at high $E_T$ may have a simpler origin: the largest transverse
energies are achieved through fluctuations rather than by a change of the
collision geometry, and, as anticipated in
Refs.~\cite{BLAIZ-OLLI,qm96}, this is
enough to produce an extra suppression with the right order of magnitude.

The model  developed in Ref.~\cite{BLAIZ-OLLI}, on which the
present analysis relies, relates  the  anomalous
$J/\Psi$ suppression to the local energy density: if the energy
density at the point where the
$J/\psi$ is produced exceeds a critical value $\epsilon_c$,
the $J/\psi$ disappears.
This model was motivated by the simple observation that the local
energy density is higher in Pb--Pb collisions than in
any other system which had been studied previously ({\it e.g.} S--U
collisions), even though the average
density in Pb--Pb collisions does not exceed that in other systems. This
simple picture was shown to account quantitatively for the data with a
single parameter, $\epsilon_c$.
Here, we extend this model taking into account transverse energy
fluctuations, which were neglected in the numerical estimates
presented in \cite{BLAIZ-OLLI}.

The transverse energy produced in a nucleus-nucleus collision can be
used as a measure of the impact parameter, the largest values of $E_T$
corresponding to the most violent collisions at small impact parameter.
However the
correlation between ${\bf b}$ and $E_T$ is not one-to-one: for a given
impact parameter, the produced transverse energy fluctuates. The effect
of these fluctuations is particularly visible for the collisions
producing the largest transverse energies: these correspond
essentially to central collisions with nearly vanishing impact parameters.
If one assumes that a fluctuation  in $E_T$ results from a fluctuation in
the energy density over the entire collision zone,  one
sees that an $E_T$ fluctuation is accompanied by an increase of the
region where the energy density exceeds $\epsilon_c$ and hence  can be
responsible for an amplification of the
$J/\psi$ suppression.

As a first estimate of this effect, let us consider central collisions of
two identical nuclei described by sharp  sphere
  densities, and ignore the nuclear absorption.
At zero impact parameter, the energy density at point ${\bf s}$ in the
transverse plane
(see Fig.\ref{F:central})  is proportional to the number of participants
in  a small tube centered at ${\bf s}$  (see below), that is
  $ \epsilon({\bf s})\propto\sqrt{R^2
-{\mathbf s}^2}$. We assume that all the $J/\psi$'s emerging at points $s <
s_c$ where the energy density is $\epsilon({\bf s})>\epsilon_c$ are
suppressed.  The radius $s_c$ of the total suppression zone is then related
to
$\epsilon_c$  by:
\begin{equation}
\frac{\epsilon_c}{\epsilon_{max}} = \sqrt{1-\frac{s_c^2}{R^2}},
\end{equation}
where $\epsilon_{max}=\epsilon(s=0)$.
The
density  of
$c
\bar{c}$ pairs produced at point ${\bf s}$ is proportional to the number of
nucleon-nucleon collisions  in a small tube
centered at
$\mathbf s$, i.e. to $ R^2-{\mathbf
s}^2$.
The survival probability follows from a simple geometrical counting:
\begin{equation}
\label{E:rpsi(0)}
\frac{\int_{s_c}^{R} d^2 {\mathbf s} \ (R^2 - s^2)}{\int_{0}^{R} d^2
{\mathbf
s} \ (R^2 -
s^2)}=\frac{(R^2-s_c^2)^2}{R^4}=
\left(\frac{\epsilon_c}{\epsilon_{max}}\right)^4.
\end{equation}
This result \cite{qm96}  provides a simple
geometrical
interpretation of the effect of a transverse energy fluctuation: assuming
such a fluctuation to be
evenly distributed over the whole transverse plane, an increase of $E_T$
results in an increase
of the volume in which the $J/\psi$'s are suppressed. At the same time,
the number of $c\bar c $ pairs, proportional to the number of
nucleon-nucleon collisions, remains constant, independent of $E_T$. This is
the origin,  in this picture, of the expected extra suppression at large
$E_T$. A fluctuation of $E_T$ by about 10\% increases
$\epsilon_{max}$ by  the same amount, and, according to
Eq.(\ref{E:rpsi(0)}), decreases the $J/\psi$ production  by  about
30\%; this is indeed  the order of magnitude of the observed effect.

\begin{figure}
\begin{center}
\epsfig{file=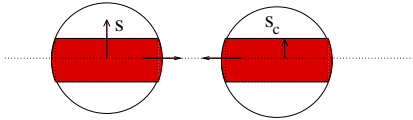,width=5.5cm}
\end{center}
\caption{Geometry of a central collision: the shaded area denotes 
the region where anomalous suppression takes place.}
\label{F:central}
\end{figure}

We now turn to a more
quantitative calculation. We assume that the transverse energy produced in
a nucleus-nucleus (A--B) collision is proportional to the number of
participants   at impact parameter $b$,  $N_p(b)$, as predicted 
in the wounded nucleon model~\cite{bialas}. 
The actual dependence may be somewhat stronger than a simple linear
relation~\cite{NA49,WA98}, 
and such deviations are important in comover scenarios
\cite{Hadronic}, 
but we do not expect them to alter our results significantly. 
Furthermore, as in
\cite{BLAIZ-OLLI} we make this relation local, and take the energy
density at point
${\mathbf s}$ and impact parameter $\mathbf b$ to be
proportional to the density $n_p({\mathbf
s},{\mathbf b})$ of participants in a plane orthogonal to the
collision axis:
\begin{eqnarray}
\label{n_part}
\lefteqn{
n_p\left({\mathbf s},{\mathbf b}\right)
= T_A\left({\mathbf s}\right) \left[1-e^{-\sigma_N T_B\left({\mathbf
b}-{\mathbf s}\right)}\right]
}\cr\ &  & \hspace*{1.5cm}
+ T_B\left({\mathbf b}-{\mathbf s}\right) \left[1-e^{-\sigma_N
T_A\left({\mathbf s}\right)}\right],
\end{eqnarray}
where $\sigma_N=32$ mb is the nucleon-nucleon inelastic cross section, and
$T_A({\mathbf s})=\int_{-\infty}^{+\infty} dz \ \rho_A({\mathbf s},z)$.
The total number of participants at impact parameter $b$ is $N_p(b)=\int
d^2{\mathbf s}
\ n_p\left({\mathbf  s},{\mathbf b}\right)$. For our numerical estimates, we
parametrize
$\rho_A({\mathbf r})$ by:
\begin{equation}
\label{E:rho}
\rho_A \left({\mathbf r}\right)=
  \frac{\rho_0}{1+\exp\left(\frac{r-R_A}{a}\right)},\qquad 
\int \rho_A({\mathbf  r}) d^3{\mathbf r}=A,
\end{equation}
with $a=0.53$ fm, $R_A=1.1 \ A^{1/3} = 6.52$ fm and $\rho_0= 0.17$
fm$^{-3}$ for $^{208}$Pb.

At fixed impact parameter, the number of participants may fluctuate. We
shall ignore these fluctuations, focussing here only on the fluctuations
of the  transverse energy produced by a fixed number of participants  or,
equivalently, at a fixed impact parameter. The corresponding
distribution is chosen to be the following gaussian
distribution \cite{KHARZ}:
\begin{equation}
\label{E:distrib}
P\left(E_T|{\mathbf b} \right)=\frac{1}{\sqrt{2\pi q^2 a N_p\left({\mathbf
b}\right)}}
\exp\left[ -\frac{\left(E_T-q N_p\left({\mathbf b}\right)\right)^2}{2q^2 a
N_p\left({\mathbf b}\right)} \right],
\end{equation}
The mean value of the transverse energy is $\langle E_T \rangle
({\mathbf b}) = q  N_p({\mathbf b})$, with $q$ the average transverse
energy per  participant, and the dispersion is $\sigma^2_{E_T}=a q^2
N_p({\mathbf b})$,  with $a$ a dimensionless parameter.
The  fit to the minimum bias $E_T$--distribution by NA50
yields
$q=0.274$~GeV and $a=1.27$ \cite{Chaurand}. Then the ``knee'' of the
$E_T$-distribution,  defined as $E_T^{knee}=q
N_p(0)$, sits at about 108 GeV.

  We
assume, as in the simple model discussed above,   that the fluctuation in
$E_T$ at given
$b$ is  distributed over the entire collision zone proportionally to
$n_p({\mathbf  s},{\mathbf b})$ so that the energy density can be written
as:
\begin{equation}
\label{E:non_saturation}
\epsilon \propto \frac{E_T}{\langle E_T \rangle ({\mathbf b})}
n_p({\mathbf s},{\mathbf b}),
\end{equation}
where $\langle E_T \rangle (b)\propto N_p(b)$ is the average
transverse energy.

The production of $c\bar c$ pairs in an A-B collision goes like the number
of nucleon--nucleon collisions, and so does   that of Drell--Yan  pairs. Thus
the probability $P(DY|b)$ that a Drell--Yan pair  be produced given that a
nucleus-nucleus collision has taken place at impact parameter
${\bf b}$ is given by:
\begin{eqnarray}
\label{E:DY_production1}
P(MB|b)\,P(DY|b)  &=&
\sigma_{DY}^{NN}
\int d^2 {\mathbf s}\,
T_A({\mathbf s}) T_B({\mathbf s}-{\mathbf b})\nonumber\\
&=&\sigma_{DY}^{NN}\ T_{AB}({\mathbf b}),
\end{eqnarray}
where $P(MB|b)=1-P_0(b)=1-\left(1-\sigma_N \frac{T_{AB}(b)}{AB}
\right)^{AB}
  \simeq 1-e^{-\sigma_N T_{AB}(b)}$ is the probability to have at least one
inelastic collision (minimum
bias) at impact parameter $b$. For large impact parameters, $P(DY|b)$
becomes
independent of $b$,
$P(DY|b)\to
\sigma_{DY}^{NN}/\sigma_N$, which is the probability to produce a
Drell--Yan pair in an inelastic
nucleon--nucleon collision.

Note that for a fixed impact parameter, the number of nucleon--nucleon
collisions may also fluctuate. We shall ignore such fluctuations,
recognizing that this is a source of uncertainty on the $E_T$ distribution
of the Drell--Yan pair production. With this simplification, the number of
Drell--Yan pairs is entirely determined by the nuclear geometry, i.e., by
the  impact parameter of the collision. Besides, the production of a
Drell--Yan pair is not affected by final state interactions, thus 
it is not directly sensitive to the transverse energy produced in the
collision  (it will become indirectly related to
$E_T$ through the relation between
$b$ and
$E_T$).

Final state interactions, however, strongly modify the $J/\psi$
production cross section. 
In addition to the standard nuclear absorption, we model the 
anomalous suppression by simply assuming~\cite{BLAIZ-OLLI} 
that any $J/\psi$ emerging  at point
$\mathbf s$ is suppressed if the energy density $\epsilon (\mathbf
s)$ is higher than
$\epsilon_c$.  Then, the probability to produce a
$J/\psi$ given that a  collision has taken place at  impact parameter $b$
and has produced a transverse energy
$E_T$ is written as follows:
\begin{eqnarray}
\label{E:psi_production1}
\lefteqn{
P(MB|b)\, P(\psi|E_T,b)  =
\sigma_{\psi}^{NN}
\int d^2 {\mathbf s}\,
\frac{1}{\sigma_a^2}
\left(1-e^{- \sigma_a T_A({\mathbf s})} \right)
}\nonumber\\ & &
\times\,\left(1-e^{- \sigma_a T_B({\mathbf s}-{\mathbf b})}
\right)
\Theta\left(n_c - \frac{E_T}{\left<E_T\right>\left({\mathbf b}\right)}
n_p\left({\mathbf s},{\mathbf b}\right) \right),
\end{eqnarray}
where $n_c$ is a parameter proportional to the critical  energy density
$\epsilon_c$ defined above.  In this formula,  $\sigma_a \approx 6.4$ mb is
the absorption cross section
\cite{NA50-2000}, and
$\sigma_{\psi}^{NN}$ is the $J/\psi$ production cross section in  a
nucleon--nucleon collision.

Putting everything together, we can write the probability to
produce a
$J/\psi$ in a  collision at a given $E_T$ as:
\begin{equation}
\label{E:psi_production2}
\frac{\int d^2 {\mathbf b} \ P(MB|{\mathbf b}) \ P(\psi|E_T,{\mathbf b}) \
P(E_T|{\mathbf b}) }
{\int d^2 {\mathbf b} \ P(MB|{\mathbf b}) \ P(E_T|{\mathbf b})}=\frac{d
\sigma_\psi / d E_T}{d \sigma_{MB}
/ d E_T}.
\end{equation}
By replacing in this formula $\ P(\psi|E_T,{\mathbf b})$ by $\
P(DY|{\mathbf b})$, we obtain the
probability to produce a Drell--Yan pair at a given $E_T$. We can then
construct the ratio:
\begin{equation}
\label{E:ratio}
{\cal N}(E_T) = \left( \frac{d \sigma_{\psi}/dE_T}{d \sigma_{MB}/dE_T}
\right)
\left/ \left(\frac{d \sigma_{DY}/dE_T}{d \sigma_{MB}/dE_T} \right)
\right. ,
\end{equation}
which we shall use in order to compare the model with experimental data.

\begin{figure}[h]
\begin{center}
\includegraphics[height=8.3cm]{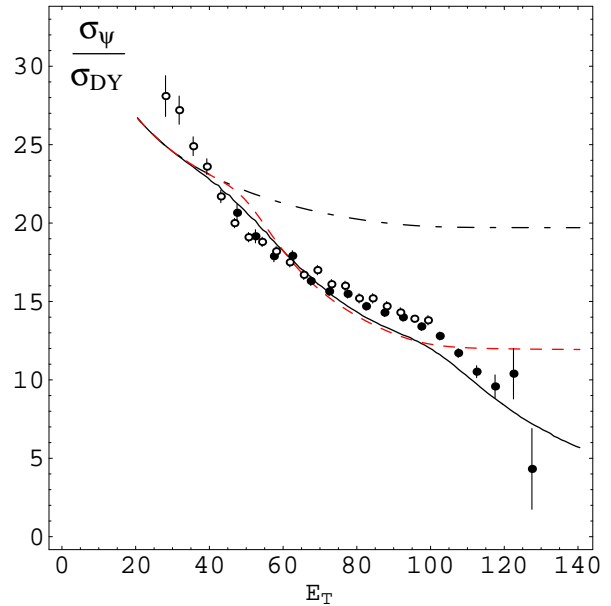}
\caption{The $J/\psi$ survival probability in a Pb--Pb
collision as a function of the transverse energy in GeV,
measured by NA50 in 1996 (open circles) and in 1998 (close circles);
after absorption in nuclear 
matter alone (dot-dashes); and after dissolution in a
quark-gluon plasma with a critical density $n_c=3.7$
fm$^{-2}$, without (dashes) and with fluctuations (full
curve).}
\label{F:suppression1}
\end{center}
\end{figure}

\begin{figure}[h]
\begin{center}
\includegraphics[height=8.3cm]{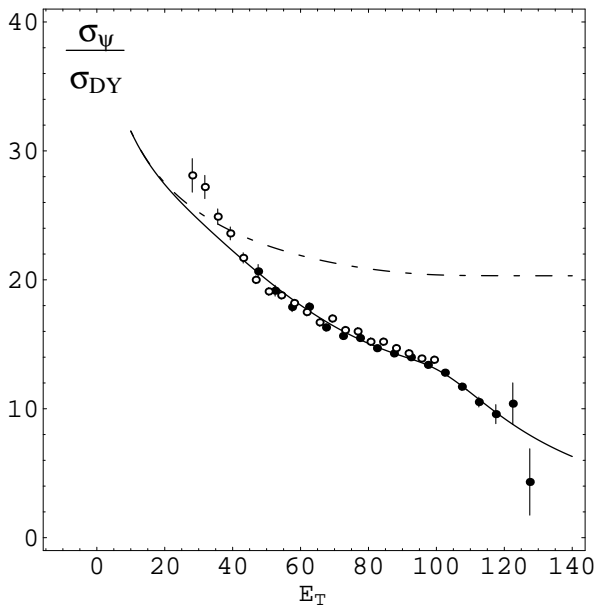}
\caption{Same as previous figure, with a gradual onset of anomalous
$J/\psi$ suppression (see text).}
\label{F:suppression2}
\end{center}
\end{figure}

Indeed, the NA50 collaboration \cite{NA50-1999} presents its results
essentially as the  same ratio, where  the numerator is an experimentally
measured quantity, while the  denominator, which is model dependent,
is deduced from a
theoretical analysis that is identical to the one presented above. 
In order to complete our theoretical estimate, we need an extra input,
the  normalisation factor
$\sigma^{NN}_{\psi}/\sigma^{NN}_{DY}$
(multiplied by the branching ratio to the dimuon decay channel)
which is obtained from experiments with lighter projectiles
and targets \cite{na38}. 
The value of this normalization factor given by NA50 is
$\sigma^{NN}_{\psi}/\sigma^{NN}_{DY}=53.5\pm 3$ \cite{Chaurand}. The
fits in Figs.\ref{F:suppression1} and \ref{F:suppression2} have
been obtained with the values 52 and 53.5 respectively.

A plot of the calculated ${\cal N} (E_T)$ is shown
in Fig.\ref{F:suppression1} together with data from NA50.
A reasonable fit of the 1998 data is obtained with a single parameter 
$n_c=3.7~\mbox{ fm}^{-2}$. Above 40~GeV, the curve deviates 
from nuclear absorption because the anomalous suppression mechanism
sets in. 
Note that the points at low $E_T$ from 1996 data \cite{NA50-1999}, 
on the other hand, are above the curve, i.e. they show less  
$J/\psi$ absorption than expected from an extrapolation of 
proton--nucleus and nucleus--nucleus data, a fact still unexplained. 
Above 100~GeV, i.e. approximately at the position of the knee, 
a second drop occurs which, as explained above,  
is associated with the increase of $E_T$ due to fluctuations.
The fact that we reproduce quantitatively the relative 
magnitude of this drop suggests, according to the discussion 
following Eq.(2), that all the $J/\psi$ mesons disappear 
at the highest densities achieved in the system.

We have to point out that the rapidity window used to mesure the $J/\psi$
and the one used to determine the $E_T$-distribution are different.
Indeed, muons
pairs are selected for $2.82 < y < 3.92$, whereas the neutral transverse
energy is
measured in the $1.1-2.3$ pseudo-rapidity window \cite{NA50-1999}. 
In the calculation presented above, we have implicitly assumed 
that the fluctuations in these two rapidity windows are 
strongly correlated. 

If, on the contrary, there were no correlation between $E_T$
fluctuations and the energy density at the point where the $c\bar c$
pair is produced, then the suppression criterion 
should involve only the average transverse 
energy at a given impact parameter, and (\ref{E:non_saturation}) 
should be replaced by $n_p({\bf s},{\bf b})>n_c$. 
The resulting ${\cal N}(E_T)$ is shown as the dashed curve in
Fig.\ref{F:suppression1}.
In this scenario, the suppression is a function of the impact
parameter {\bf b} only: as a consequence, the suppression 
saturates at large $E_T$, to its value at $b=0$.
Such a model for $J/\Psi$ suppression is proposed
in \cite{Nardi97,QGP}, where,
indeed, the criterion for suppression depends only on {\bf b}.
Although it refers to the density of strings rather than to that of 
participants, this bears little influence on the results.

Clearly, our theoretical curve in Fig.\ref{F:suppression1}
does not provide a perfect fit to NA50 data,
which have small error bars. 
In order to improve the quality of the fit, we did another
calculation with a more gradual suppression mechanism: 
namely, we replaced the $\Theta$--function in
Eq.(\ref{E:psi_production1}) by $[1-\tanh \lambda(n-n_c)]/2$. 
The results for $n_c=3.75~{\rm fm}^{-2}$ and $\lambda=2~{\rm fm}^2$
are displayed in Fig.\ref{F:suppression2}. A perfect fit of the 1998
data is obtained over the entire $E_T$ range from 40 GeV to 120
GeV. Note that above 80~GeV, the suppression becomes 
total at the hottest point of the system, 
so that the behaviour at large $E_T$ is essentially the same as 
in Fig.\ref{F:suppression1}. 
We obtain a fit indistinguishable from that of 
Fig.\ref{F:suppression2} with a two-threshold scenario \cite{NA50-2000}, 
by assuming that 40\% of 
the $J/\psi$ are suppressed
if the density lies between $n_{c1}=3.3~{\rm fm}^{-2}$ and 
$n_{c2}=4.0~{\rm fm}^{-2}$ and 100\%  above $n_{c2}$. 
Note that the second threshold does not produce any visible structure in 
the $E_T$ dependence of $\sigma_\psi/\sigma_{\rm DY}$.

Finally, we would like to compare our model with that recently
proposed in \cite{CAPEL}, where
the $J/\Psi$ suppression is explained by
interactions with comoving particles. The density of
comovers is related to the transverse energy in the same way
as the energy density in our model.
Thus these authors also obtain increased $J/\Psi$ suppression, 
but much less pronounced than in NA50 data.
This is because, as explained in detail in \cite{qm96},  
absorption by comovers depends more weakly on tranverse 
energy fluctuations than in the present model.

In summary, we have shown that the second drop in the 
$J/\psi$ yield around the knee of the $E_T$--distribution
can be interpreted as an effect of $E_T$ fluctuations, 
if one assumes that $J/\psi$ suppression is related to the local energy
density in the system. 
We have achieved our best fit by assuming that the suppression increases 
gradually with the energy density; in particular, 
there is no indication in the data that the suppression occurs in two steps. 
However, in order to reproduce the magnitude of the observed ``second drop'',
we need to 
assume that all the $J/\psi$'s disappear at the highest densities achieved 
in a Pb--Pb collision. 

\acknowledgements

We thank A. Capella and D. Kharzeev for discussions 
and B. Chaurand for detailed
explanations concerning the analysis done by the NA50 Collaboration.

\end{document}